\documentclass[fleqn,usenatbib]{mnras}

\usepackage{newtxtext,newtxmath}
\usepackage[T1]{fontenc}

\DeclareRobustCommand{\VAN}[3]{#2}
\let\VANthebibliography\thebibliography
\def\thebibliography{\DeclareRobustCommand{\VAN}[3]{##3}\VANthebibliography}

\usepackage{graphicx}
\usepackage{amsmath}
\usepackage{caption}
\usepackage{subcaption}
\usepackage{gensymb}
\usepackage{comment}

\title[A New Superbubble Finding Algorithm]{A New Superbubble Finding Algorithm: Description and Testing}

\author[B. Wallin et al.]{
Brock Wallin,$^{1}$ 
Benjamin D. Wibking,$^{1}$\thanks{Corresponding author: benwibking+astro@gmail.com}
 and G. Mark Voit$^{1}$
\\
$^{1}$Department of Physics and Astronomy, Michigan State University, 567 Wilson Rd, East Lansing 48824, Michigan, USA
}

\date{Accepted XXX. Received YYY; in original form ZZZ}

\pubyear{\the\year{}}

\begin{document}
\label{firstpage}
\pagerange{\pageref{firstpage}--\pageref{lastpage}}
\maketitle

\begin{abstract}
We present a new algorithm for identifying superbubbles in H~I column density maps of both observed and simulated galaxies that has only a single adjustable parameter. The algorithm includes an automated galaxy-background separation step to focus the analysis on the galactic disk. To test the algorithm, we compare the superbubbles it finds in a simulated galactic disk with the ones it finds in 21~cm observations of a similar galactic disk. The sizes and radial distribution of those superbubbles are indeed qualitatively similar. However, superbubbles in the simulated galactic disk have lower central H~I column densities. The H~I superbubbles in the simulated disk are spatially associated with pockets of hot gas. We conclude that the algorithm is a promising method for systematically identifying and characterizing superbubbles using only H~I column density maps that will enable standardized tests of stellar feedback models used in galaxy simulations.
\end{abstract}



\section{Introduction} \label{sec:intro}

Spatially correlated supernovae can excavate large cavities called \textit{superbubbles} within galactic disks, recognizable as hot, low-density, ionized regions surrounded by higher density shells of swept up gas \citep[e.g.,][]{MacLow_1988}. Once gas in a galactic disk becomes self-gravitating, it gives birth to massive stars. Strong stellar winds from those massive stars then blow bubbles filled with hot $\left(T\gtrsim10^6\ \mathrm{K}\right)$ gas \citep{Weaver_1977}. Shortly thereafter, the massive stars end their lives with supernova explosions \citep[e.g.,][]{Maoz_Review_2014} that locally disrupt star formation \citep[e.g.,][]{Tacconi_Review_2020}. When multiple stars in the same cluster go supernova within a few tens of millions of years of each other, their supernova remnants combine to increase the size of the bubble by adding energy to the bubble's interior. Together, clustered supernovae and strong stellar winds can thereby produce superbubbles hundreds of parsecs in size, capable of blowing through the top and bottom of a galactic disk, producing kpc-scale holes $\mathrm{HI}$ holes in it, and allowing the hot gas interior to the bubble to erupt into the circumgalactic medium (CGM) \citep[e.g.,][]{Tomisaka_1986, Brinks_1986, MacLow_1988}. 

Several recent analyses of supernova feedback and its impact on galaxy evolution have focused attention on the importance of superbubbles and how their properties depend on correlated supernovae \citep[e.g.,][]{Orr_2022, Oku_2023, Vasiliev_2023, Tan_2024, Li_2024}. This paper is aimed at understanding whether numerical simulations of galaxy evolution produce superbubbles with properties similar to those in real galactic disks. As an example, we look at the superbubbles created by the magneto-hydrodynamic (MHD) simulation of \cite{Wibking_2022} and compare them to the superbubbles found in a 21-cm radio observation of the face-on disk galaxy NGC 6946 \citep{Walter_2008}.  We know of few such comparisons between simulated and observed superbubbles in the recent literature. This work is intended to bridge that gap. 

The simulated galaxy is Milky Way-like and its stellar feedback processes include both supernovae and ionizing photons from young stars. Star particles created in the simulation produce supernovae at a constant rate of $3 \times 10^{-4}\ \mathrm{SNe}\ \mathrm{M_\odot}^{-1}\ \mathrm{Myr}^{-1}$ at an age within $0<t_{\mathrm{age}}<30\ \mathrm{Myr}$. The sizes and locations of the superbubbles they produce depend on the spatial and temporal correlations of star formation in the simulated galaxy.

To identify those superbubbles, we developed a simple and unbiased algorithm that can easily be applied to both simulations and observations of galactic disks. Several algorithms have previously been developed to automatically identify and measure interstellar filaments and bubbles in observed \citep[e.g.,][]{Daigle_2003, Men'shchikov_2010, Men'shchikov_2012, Collischon_2021} and simulated \citep[e.g.,][]{Williams_1994} galaxies. \cite{Boomsma_2008} performed an analysis of NGC 6946 and identified H~I holes and bubbles using 21~cm column-density and velocity information. Similar analyses have been done for other galaxies, such as NGC 2403 \citep{Thilker_1998}, IC 2574 \citep{Walter_1999}, and IC 10 \citep{Wilcots_1998}. 

The algorithm presented here is simpler. It is designed to identify likely superbubble candidates using just an H~I column density map with no velocity dispersion information. Simplicity was a key goal, so that human input could be kept to a minimum, thereby reducing confirmation bias in determining the superbubbles. The algorithm also determines the sizes and spatial distributions of the superbubbles, so that simulations and observations can be directly compared. 

The rest of this paper is organized as follows: Section \ref{sec:dataset} describes the simulated and observed galaxies used in the development and testing of the algorithm. Section \ref{sec:method} presents the superbubble finding algorithm in its entirety, from an input column density map to an output list of identified superbubbles and their properties. In section \ref{sec:result}, we show the results of applying the algorithm to both the simulated galaxy and NGC 6946. Section \ref{sec:discuss} discusses the performance of the algorithm and suggests steps for future work to test and improve it. Section \ref{sec:conclusion} summarizes our work.


\section{Data Set} \label{sec:dataset}

\begin{figure}
\includegraphics[width=\columnwidth]{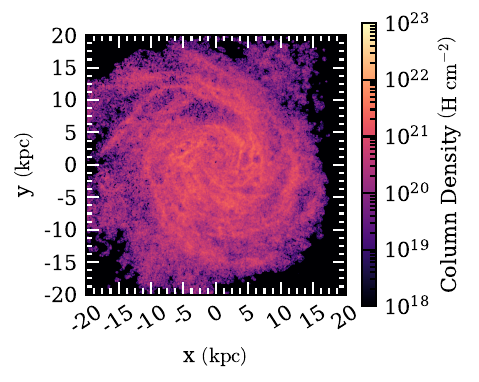}
\caption{Neutral hydrogen column density map of NGC 6946 using data acquired from the THINGS database \citep{Walter_2008}.
\label{fig:NGC6946_im}}
\end{figure}

\begin{figure}
\includegraphics[width=\columnwidth]{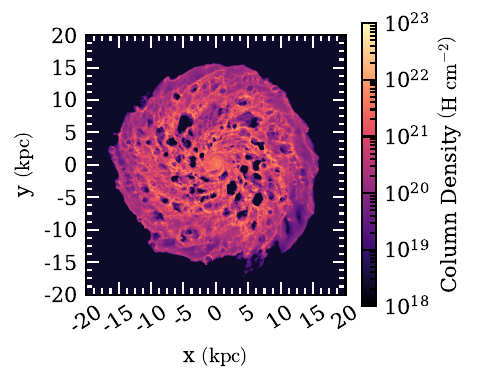}
\caption{Neutral hydrogen column density map of a face-on simulated galaxy from \citet{Wibking_2022}.
\label{fig:sim_im}}
\end{figure}

For development and testing of the algorithm, we used a 21-cm observation of NGC 6946 \citep{Walter_2008} and a simulated galaxy from \cite{Wibking_2022}. Figures \ref{fig:NGC6946_im} and \ref{fig:sim_im} show H~I column density maps for these galaxies. Superbubbles from supernova explosions can be seen in both images as regions within the H~I disk that have lower column density than the surrounding gas.

\subsection{NGC 6946} \label{subsec:NGC6946}

We selected NGC 6946 for comparison with the simulated galaxy because its physical size is similar. The star formation rate of NGC 6946 is reported to be $4.76\ \rm M_\odot\ yr^{-1}$, and it has an H~I mass of $4.15\times 10^9\ \rm M_\odot$ \citep{Walter_2008}. The superbubbles in NGC 6946 appear as fuzzy regions not much lower in column density than the rest of the galactic disk, making them challenging to identify by eye. Overcoming that challenge is one of the main goals of the algorithm. The simulated column-density map in Figure \ref{fig:sim_im} is sharper because it does not include the noise and beam effects associated with the radio observation. We would like to determine superbubble properties that are independent of those effects.

Our algorithm takes as input H~I column density maps that have pixel values in units of $\mathrm{H}\ \mathrm{cm}^{-2}$, so we first need to convert both the simulated galaxy and NGC 6946 data files into those units. Starting with NGC 6946, the FITS tables provide velocity-integrated H~I 21~cm intensities ($S \Delta v$) in units of $\mathrm{Jy}\ \mathrm{beam}^{-1}\ \mathrm{m}\ \mathrm{s}^{-1}$. To convert those intensities into velocity-integrated surface brightness temperatures, we account for the beam size using the formula 
\begin{equation}
    T_{\rm B} \Delta v 
        = \frac {(6.07 \times 10^5) S \Delta v}
                {\mathrm{FWHM}_{\mathrm{maj}} \times \mathrm{FWHM}_{\mathrm{min}}}            
    \label{eq:Walter_1}
\end{equation}
which gives $T_{\rm B} \Delta v$ in units of ${\rm K \, km \, s^{-1}}$ when $S\Delta v$ is in ${\rm Jy \, beam^{-1} \, km \, s^{-1}}$ and the full width at half max of the beam's major axis (FWHM$_{\rm maj}$) and minor axis (FWHM$_{\rm min}$) are in units of arseconds \citep{Walter_2008}.
The H~I column density is then
\begin{equation}
    N_{\mathrm{H \, I}}=1.823\times 10^{18}\ T_{B}\ \Delta v
\label{eq:Walter_5}
\end{equation}
Finally, the outer regions of the NGC 6946 21~cm map can have unphysically small and sometimes negative flux values. Those are set to $10^{18}\ \mathrm{H}\ \mathrm{cm}^{-2}$.

\subsection{Simulated Galaxy} \label{subsec:sim_gal}

Superbubbles in the simulated galaxy are more obvious than in NGC 6946 because the minimum H~I column densities near their centers are much smaller. The simulated galaxy is evolved for $570\ \rm Myr$ with gas particles of mass $859.3\ \rm M_\odot$ and total H~I gas mass of $4.06 \times 10^9\ \rm M_\odot$, and at the time of observation has a star formation rate of ${\sim}2\ \rm M_\odot\ yr^{-1}$ \citep{Wibking_2022}.

Our pre-processing procedure to create a synthetic H~I map from the simulation follows these steps:

\begin{description} 

  \item[\textbf{Step 1 -- Density Conversion:}] The total gas density in the simulation is converted to hydrogen density using a hydrogen to total gas mass fraction of $X_{\mathrm{H}} = 0.71$. Any unphysical gas temperature and density values are adjusted: temperature values in very low temperature regions are set to $1\ \mathrm{K}$, and density values in very low density regions are set to $10^{-6}\ \mathrm{g}\ \mathrm{cm}^{-3}$. We also use temperature information to mask all density particles with temperatures greater than or equal to $10^4\ \mathrm{K}$, in order to remove all ionized gas from the simulation. 
  
  \item[\textbf{Step 2 -- Dust Optical Depth:}] The next few steps account for the ratio of neutral atomic hydrogen to molecular hydrogen, following \citet{Krumholz_2011} in its entirety. That ratio depends on the disk's optical depth. To determine it, we estimate the simulated disk's scale height using the formula $h=\rho /|\nabla \rho|$, 
  where $\rho$ is the disk's gas density evaluated at each cell, and $|\nabla \rho|$ is the magnitude of the density gradient, with the gradient calculated between all neighboring cells across the full grid. At times the gradient is very small which makes the scale height very large. To account for this, we set the maximum scale height of the simulated galaxy to be $100\ \mathrm{pc}$, approximately the scale height of the Milky Way's disk. This is done because large differences in scale height will have a large impact on the final column density values. The disk's column density $\Sigma=\rho h$ is then calculated and used to find the dust optical depth $\tau_c=\Sigma \sigma_d/\mu_\mathrm{H}$, taking $\sigma_d=\left(10^{-21}\ \mathrm{cm}^{-2}\right)\ \sigma_{d,-21}\ $ to be the dust cross section per hydrogen nucleus and $\mu_\mathrm{H}=2.3\times 10^{-24}\ \mathrm{g}$ to be the mean mass per hydrogen nucleus. 
  
  \item[\textbf{Step 3 -- Scaled Radiation Field:}] To quantify the scaled UV radiation field within the simulated disk, we start with this equation from \cite{Krumholz_2011}:
  \begin{equation}
        \chi=71\left(\frac{\sigma_{d,-21}}{\mathcal{R}_{-16.5}}\right)\frac{G_0^{'}}{n_{\mathrm{H},0}}
        \label{eq:Krumholz_3}
    \end{equation}
  Here, $\sigma_{d,-21}$ represents the dust cross section per hydrogen nucleus to $1000\ \mathrm{\AA}$ photons, $\mathcal{R}_{-16.5}$ represents the rate coefficient for $\mathrm{H}_2$ formation on dust grains, normalized to a Milky Way value of $10^{-16.5}\ \mathrm{cm}^{3}\ \mathrm{s}^{-1}$, $G_0^{'}$ represents the ambient UV radiation field intensity normalized to $2.72\times 10^{-3}\ \mathrm{ergs}\ \mathrm{cm}^{-3}\ \mathrm{\AA}^{-1}$ \citep{Parravano_2003}, and $n_{\mathrm{H},0}$ represents the number density of hydrogen nuclei. We assume that $\left(\sigma_{d,-21}/\mathcal{R}_{-16.5}\right)=1$ and also that the ambient UV radiation field intensity has a constant value of 1, so that 
  \begin{equation}
    \chi=\frac{71}{n_{\mathrm{H},0}}
    \label{eq:chi}
  \end{equation}
  
  \item[\textbf{Step 4 -- H~I Fraction:}] Using the dust optical depth and scaled radiation field from steps 3 and 4 we compute the $s$ parameter: 
  \begin{equation}
    s=\frac{\ln{(1+0.6\chi+0.01\chi^{2})}}{0.6\tau_c}
    \label{eq:Krumholz_2}
  \end{equation}
  This parameter is used by \cite{Krumholz_2011} to determine the fraction of neutral atomic hydrogen, via the fitting formula
  \begin{equation}
    f_{\mathrm{HI}}\simeq\left(\frac{3}{4}\right)\frac{s}{1+0.25s}
    \label{eq:Krumholz_1}
  \end{equation}
  We set $f_{\rm HI}$ equal to unity if this formula gives a greater value.
  
  \item[\textbf{Step 5 -- H~I Column Density:}] Using the H~I fraction calculated in the previous step, we find the projected column density map of the simulated galaxy using a $33\degree$ inclination to match the observation of NGC 6946 \citep{Walter_2008}. We sum the H~I density of the simulated galaxy along the line of sight at $33\degree$ to the plane of the galaxy to get a column density map in units of $\mathrm{H}\ \mathrm{cm}^{-2}$. 
  
  \item[\textbf{Step 6 -- Beam Smoothing:}]  With the simulation data in the form of a 21-cm H~I column density map, we follow these steps to create a synthetic observation. We create an empty 2D array with the same size as the column density map and fill the array with a 2D Gaussian, centered at the central pixel. The 2D Gaussian array is created with a mean of $0$ and a standard deviation of $\sigma=6.04\ \mathrm{arcsec}$ in both the x and y directions. This standard deviation is the $\mathrm{FWHM}_{\mathrm{maj}}$ of the beam used when observing NGC 6946. To combine the H~I column density map and the 2D Gaussian array, we first take the 2D fast Fourier transform of each. Next, both transformed arrays are multiplied together element-wise, producing a 2D map of the combined column density and 2D Gaussian in Fourier space.
  
  \item[\textbf{Step 7 -- Adding Noise:}]  To construct the synthetic 2D noise array, we again create an empty 2D array with the same size as the H~I column density map. The synthetic noise is sampled from a 1D Gaussian with a mean of $0$ and standard deviation found as follows. We take the 2D fast Fourier transform of the NGC 6946 H~I map, then calculate the standard deviation of the transformed map and use this value as the standard deviation of the 1D Gaussian. Then we randomly sample from the 1D Gaussian for each element in the empty 2D array to create a synthetic 2D noise array in Fourier space.
  \\ 
  Before adding the 2D noise array to the H~I map, we ensure the condition $\mathcal{F}\left[I(\omega)\right]=\mathcal{F}\left[I(-\omega)^{*}\right]$ holds. This says that for any frequency of the 2D array found in Fourier space, the corresponding Fourier space element of the array with this frequency must be the complex conjugate of the Fourier space element of the array corresponding to the negative frequency. Each frequency is looped through to ensure this condition holds, and where this condition does not hold we set the negative frequency element to be the complex conjugate of the positive frequency element. 
  \\
  We deal with the special conditions of the four corners and four edges separately. For each of the four corners, we take their absolute value and use this as their new value. We handle the edges minus the four corners in a similar way as to the rest of the image, but we only focus on either the $x$ frequency or $y$ frequency at a time, keeping the other the same. First, the element corresponding to the positive of the $x$ frequency is set to be the complex conjugate of the corresponding negative $x$ frequency keeping the $y$ frequency the same. We repeat this for the $y$ frequency, keeping the $x$ frequency component the same. When the Fourier transform is inverted to get a column density map in physical space, this step ensures that the noise-added map's column density values are real.
  \\
  To combine the 2D noise array with the synthetic beam applied H~I map, both 2D arrays are in Fourier space already and we add them element-wise. Then we take the 2D inverse fast Fourier transform of the combined array to get the synthetic observation we will use in the superbubble identifying algorithm.
  
\end{description}

\begin{figure}
\includegraphics[width=\columnwidth]{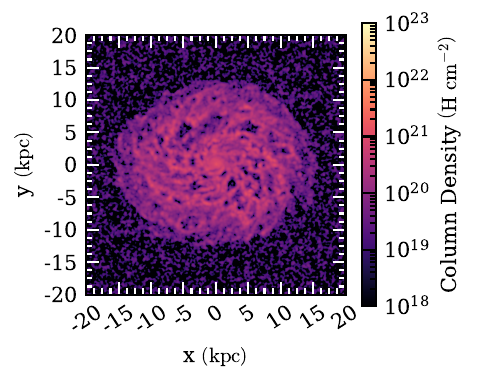}
\caption{Pre-processed H~I column density map of the simulated galaxy, inclined by 33\textdegree \ to match the inclination of NGC 6946.
\label{fig:processed_sim}}
\end{figure}

\begin{figure*}
\centering
\includegraphics{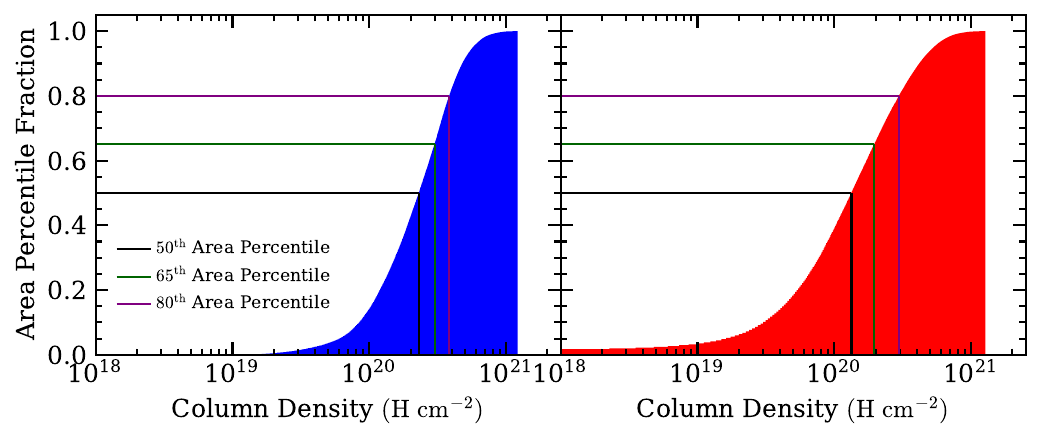}
\caption{Cumulative probability distribution functions and percentile thresholds for H~I column density. NGC6946 is on the left and the simulation is on the right. Only the galactic disks are considered; the exterior regions are ignored. Colored lines relate the $50^{\text{th}}$, $65^{\text{th}}$, and $80^{\text{th}}$ area percentiles to the column density thresholds used in this analysis. All pixels with lower column density than a threshold value are assigned to candidate superbubbles by the algorithm.
\label{fig:cumu_hist}}
\end{figure*}

\subsection{Discussion of Images} \label{subsec:more_manipulations}

The procedure used to make the simulated image entails many assumptions. To mitigate their effects, we applied Gaussian smoothing to both the NGC 6946 image and the pre-processed simulated image, so that small-scale structures are softened. The Gaussian kernel has a standard deviation corresponding to twice the semi-major axis FWHM value of the applied beam. We use a factor of $2$ here to ensure noticeable smoothing in the image without erasing the superbubble-scale features we are trying to detect. Figure \ref{fig:processed_sim} shows the final result for the simulated galaxy and is the input image fed into the superbubble finding algorithm.

There is a noticeable difference in minimum column density values between the simulated image and NGC 6946. Typically, NGC 6946 has larger H~I column densities than the simulated image. These differences are most evident when looking at the centers of the superbubble regions, where the simulation has very low H~I column densities. In NGC 6946 the central column densities within each superbubble region are larger, making the superbubbles less evident and more difficult to pick out. 


\section{Methods} \label{sec:method}

The algorithm we have developed automatically identifies and measures H~I holes, which may be physical H~I superbubbles that have broken out of the galactic disk (see \S \ref{subsec:temp_analysis}). The algorithm was designed with as few adjustable parameters as possible to minimize any confirmation bias. Unlike other algorithms mentioned in the introduction, we avoid using 21-cm velocity information in order to keep the model simple and usable with only a column density map. Our algorithm consists of the following steps:
\begin{description}
  \item[\textbf{Step 1 (Section \ref{subsec:ext_rem}):}] Given a column density map of neutral atomic hydrogen taken from a simulation or a 21-cm emission observation, first embed the image in a square array. Then, define an exterior region around the galactic disk and remove it from the analysis, leaving just the galactic disk.
  \item[\textbf{Step 2 (Section \ref{subsec:region_identify}):}] Using the disk map produced in Step 1, find all regions that fall below a column density threshold defined by a fixed percentile of column density by area.
  \item[\textbf{Step 3 (Section \ref{subsec:bub_conditions}):}] Using the regions identified in Step 2, mask out regions that are unlikely to be superbubbles. The remaining regions are considered candidate superbubbles.
\end{description}

\subsection{Exterior Removal} \label{subsec:ext_rem}

The first step the algorithm performs is to find and remove the exterior region around the galactic disk, so that we are left an image of just the galactic disk. By automatically removing the exterior, we can perform the algorithm on any input image.

\begin{description}

\item[\textbf{Step 1a:}] The algorithm first selects an H~I  column density threshold for defining the disk. It takes the arithmetic average of the H~I column density to be the threshold cut, and then calculates the image area above that threshold. Next, it computes an approximate radius $r$ corresponding to this area, using the formula $r\approx \sqrt{A/2}$, dividing by $2$ rather than $\pi$ to obtain a more conservative estimate of the disk's size. This choice decreases the possibility that some of the galactic disk could be incorrectly masked during the exterior removal process.

\item[\textbf{Step 1b:}] Next, the algorithm smooths the image with a uniform filter that has a smoothing scale equal to the radius $r$ from Step 1a. For each pixel, this filter averages all pixels within the smoothing scale and sets the pixel's value to that average. For pixels near the edge, pixel values within the filter that would extend over the edge of the image are taken to have the value of the nearest edge pixel.

\item[\textbf{Step 1c:}] The algorithm then uses one-fourth of the maximum column density in the smoothed image from Step 1b as a threshold to cut the smoothed image into two regions. Everything below the threshold is considered exterior. Everything above the threshold is considered the galactic disk. Using one-fourth of the maximum ensures that the cut will not be affected by extremely large exterior regions and will not require fine-tuning by the user. In each case we have tested, masking the image like this leaves only two contiguous regions, the galaxy region and the exterior, which can be cleanly separated. 
\end{description}

\begin{figure*}
\includegraphics[width=16cm]{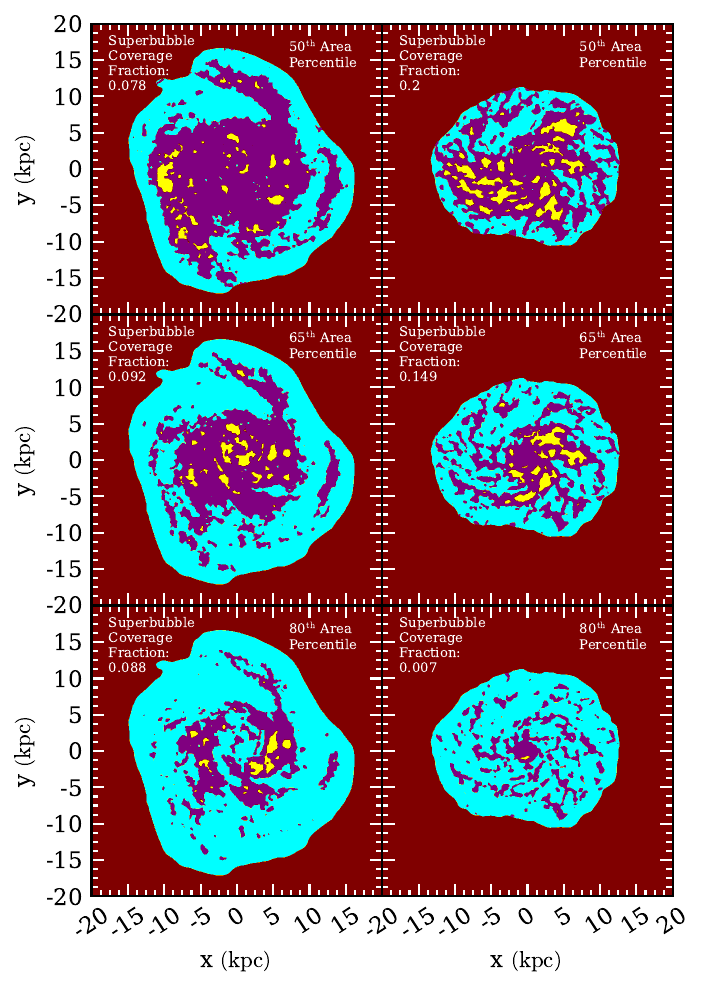}
\caption{Superbubble candidates in NGC~6946 (left column) and the simulated galaxy (right column). The column density thresholds used are the $50^{\text{th}}$ (top row), $65^{\text{th}}$ (middle row), and $80^{\text{th}}$ (bottom row) area percentiles. Yellow regions are candidate superbubbles. Purple regions are parts of the galactic disk above the column-density threshold. Aqua regions are below the threshold but excluded from consideration as superbubbles because they fail one of the exclusion criteria. Crimson regions are considered exterior to the galactic disk. Each image lists the fraction of the disk's area covered with (yellow) superbubbles.
\label{fig:bubble_im}}
\end{figure*}

\begin{figure*}
\includegraphics[width=16cm]{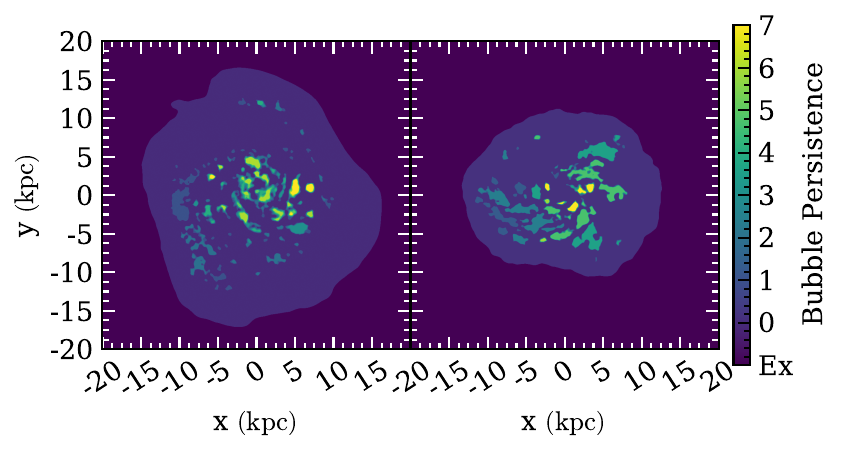}
\caption{Bubble persistence maps for NGC~6946 (left) and the simulated galaxy (right). Each map is based on applying the algorithm seven times, with percentile thresholds going from $50^{\text{th}}$ to $80^{\text{th}}$ in steps of $5\%$. Colors show the number of thresholds at which a given pixel is part of a candidate superbubble.
\label{fig:bubble_persistence}}
\end{figure*}

\subsection{Identifying Candidate Superbubbles} \label{subsec:region_identify}

After the exterior is removed, the next step is identifying candidate superbubbles using isocontours of column density.

\begin{description}
    \item[\textbf{Step 2a:}] The column density threshold for a given galaxy is chosen based on a given column density percentile $P$. The user chooses the value of $P$ for all galaxies of interest, and the algorithm calculates the corresponding column density thresholds for each galaxy. This method was chosen because of the systematic differences in typical column density values between NGC 6946 and the simulated galaxy. (Using a fixed percentile enables a direct comparison of the identified superbubbles between the two maps, even if the corresponding column density thresholds are very different.)

    \item[\textbf{Step 2b:}] Using the column density threshold calculated in Step 2a, the algorithm selects all pixels in the galaxy image whose values are below the threshold, so that $P$ percent of the disk's area is below the threshold. Contiguous sub-threshold regions are considered candidate superbubbles. The algorithm numbers each candidate and measures its area and centroid using the regionprops function from the SciPy library.
\end{description}

\subsection{Excluding Unlikely Superbubbles} \label{subsec:bub_conditions}

With the image processed and potential superbubble regions identified, the final step is to exclude regions that are unlikely to be real superbubbles. Once we remove unlikely superbubbles, we are left with regions that collectively cover a smaller area of the galactic disk than the selected percentile threshold.  We call that fractional area the ``superbubble coverage fraction" and use that number to quantify the total area of the disk identified as superbubbles.

\begin{description}
    \item[\textbf{Step 3a:}] We exclude regions smaller than a minimum superbubble size. The minimum size is set by the size of the beam used in observing the galaxy, since features smaller than the beam size are unphysical. The algorithm uses a physical size corresponding to twice the size of the beam as the lower limit of a superbubble’s approximate diameter. All regions that fulfill $\sqrt{A/\pi}<D$, where $D$ is the semi-major axis diameter of the beam, and $A$ is the total area of the superbubble, are masked out. We use twice the size of the beam rather than the beam size itself because we want to look for superbubbles that would be resolvable as extended features, rather than small scale fluctuations in the column density map that may be more likely to be caused by noise.

    \item[\textbf{Step 3b:}] We exclude regions that are contiguous with the exterior. All pixels directly bordering the exterior, all pixels bordering these pixels, and so on, are identified. (A border with a width of 5 pixels is used to ensure all regions close to the exterior are removed.) Any candidate superbubble contiguous with the border is then excluded.
\end{description}


\section{Results} \label{sec:result}

\subsection{Algorithm Output} \label{subsec:output}

Here we discuss how the algorithm's results depend on choices for the area percentile, $P$. We will focus on the $50^{\text{th}}$, $65^{\text{th}}$, and $80^{\text{th}}$ percentiles. Figure \ref{fig:cumu_hist} highlights these area percentiles on the cumulative column density histogram for both NGC 6946 and the simulated galaxy. A particular area percentile applied to both NGC 6946 and the simulated galaxy corresponds to different column density thresholds. The column density threshold values used as the cut to determine the superbubbles for NGC 6946 and the simulated galaxy for the $50^{\text{th}}$ area percentile are $10^{20.36}\ \mathrm{H}\ \mathrm{cm}^{-2}$ and $10^{20.13}\ \mathrm{H}\ \mathrm{cm}^{-2}$ respectively. By using an area percentile rather than a specific column density enables us to directly compare these two galaxies and their superbubble properties.

After applying the algorithm to the simulated galaxy and NGC 6946, we produced images of the identified superbubbles for the selected area percentile. Figure \ref{fig:bubble_im} shows the results using the $50^{\text{th}}$, $65^{\text{th}}$, and $80^{\text{th}}$ area percentiles for both galaxies. The $50^{\text{th}}$ area percentile identifies the most superbubbles for both the simulated galaxy and NGC 6946. There is a trend in both the simulated galaxy and NGC 6946: the concentration of superbubbles moves towards the center of the galactic disk and the number of superbubbles found decreases as $P$ increases. Both galaxies have higher H~I column density near their centers, so as the column density threshold increases, the identified superbubbles become more concentrated near the center of the galaxy. Increasing the area percentile also causes more of the galaxy to become below the threshold, allowing candidate superbubbles to be increasingly interconnected, contiguous with the exterior, and removed from consideration, as can be seen in Figure \ref{fig:bubble_im}.

Figure \ref{fig:bubble_im} shows all the remaining superbubble regions in yellow and shows the excluded regions in aqua. The remaining purple region is the part of the galactic disk that is above the column density threshold. After both exclusion conditions are applied, the remaining superbubble candidates are labeled and saved. The superbubble coverage fraction reported in each panel represents the ratio of (non-excluded) superbubble area (in yellow) to remaining galactic disk area (yellow, purple). It does not include the aqua parts consisting of low column density regions either contiguous with the exterior or smaller than the minimum size requirements.

In the simulated galaxy, the superbubble coverage fraction decreases as $P$ increases, but this trend does not hold for NGC 6946. There the superbubble coverage fraction decreases as $P$ goes from $65$ to $80$ to $50$. This difference arises because regions of high column density in NGC~6946 remain more contiguous as $P$ increases, allowing larger superbubbles to persist up to the $80^{\text{th}}$ area percentile.

\subsection{Bubble Persistence} \label{subsec:bub_persist}

We tested the stability of the algorithm by exploring what we call “bubble persistence.” We define bubble persistence as the number of area percentiles at which the same superbubble is found. Figure \ref{fig:bubble_persistence} provides a visual representation of bubble persistence. We take the superbubble images (yellow regions in Figure \ref{fig:bubble_im}) and add them together. We add a total of $7$ superbubble images, going from the $50^{\text{th}}$ to $80^{\text{th}}$ percentiles in steps of $5$ percentile points, and count the number of times that a particular pixel belongs to a superbubble. Pixels that qualify as ``superbubble" many times have high ``bubble persistence" and indicate where a candidate superbubble is robust. There is a trend of higher bubble persistence towards the center of the galactic disk for both galaxies, which can be attributed to the generally larger column density near the center of the galaxy.

\begin{figure}
\includegraphics[width=\columnwidth]{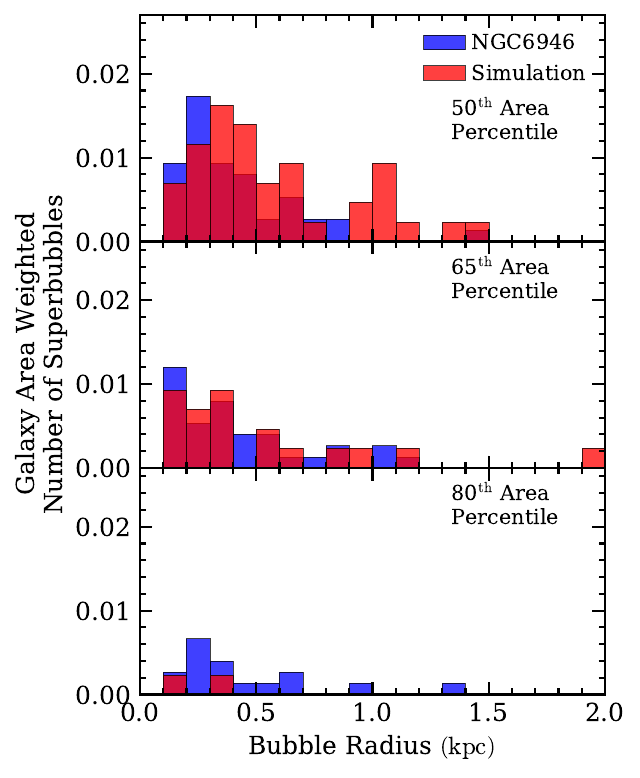}
\caption{Distribution of superbubble radius for NGC 6946 (blue) and the simulated galaxy (red) at the $50^{\text{th}}$, $65^{\text{th}}$, and $80^{\text{th}}$ area percentiles, from top to bottom.\\
\label{fig:bubble_radii}}
\end{figure}

The algorithm allows us to compare the radii of the identified superbubbles which we demonstrate by creating superbubble radius histograms for the $50^{\text{th}}$, $65^{\text{th}}$, and $80^{\text{th}}$ area percentiles using bins of $0.1\ \mathrm{kpc}$. In Figure \ref{fig:bubble_radii} we show superbubble radius histograms, weighted by the total area of each galaxy, to show how much of the galactic disk is covered with superbubbles at each radius. We see a shared trend: smaller superbubbles collectively cover more of a galaxy's disk than larger ones. The likely origin of this trend is that much more supernova energy is required to make a large bubble. Since we observe similar trends in galaxy coverage, we are encouraged that the algorithm can find superbubbles and measure their sizes consistently when given different column density maps.

\begin{figure}
\includegraphics[width=\columnwidth]{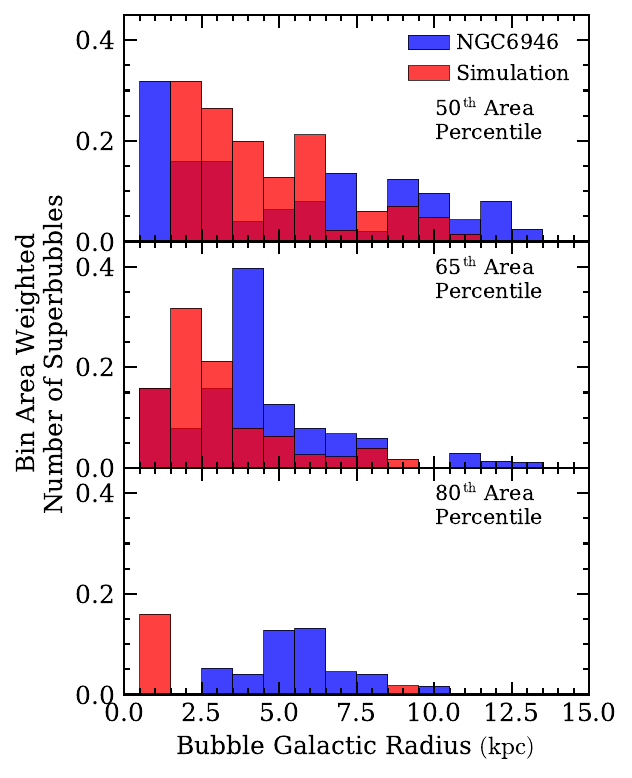}
\caption{Distribution of superbubble distance from the galactic center for NGC 6946 (blue) and the simulated galaxy (red) at the $50^{\text{th}}$, $65^{\text{th}}$, and $80^{\text{th}}$ area percentiles, from top to bottom.
\label{fig:bubble_galrad}}
\end{figure}

Figure \ref{fig:bubble_galrad} presents similar histograms showing the number of superbubbles found within different ranges of galactic radius. The number of superbubbles found in each radial bin is weighted by the area of that annulus. Each bin covers $1\ \mathrm{kpc}$ in radius from the galactic center outward. Note that the fractional number of superbubbles decreases toward larger galactic radii. 

\begin{figure}
\includegraphics[width=\columnwidth]{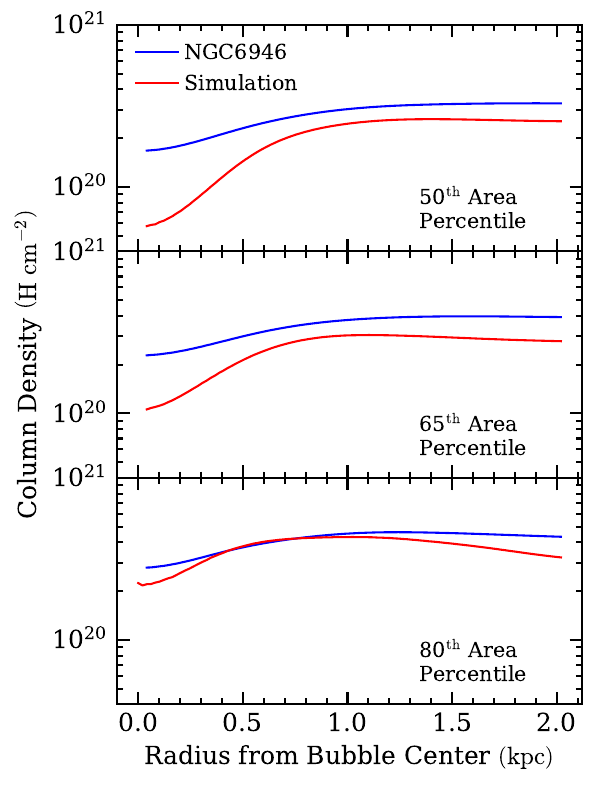}
\caption{Average superbubble column density profile for NGC6946 (blue) and the simulated galaxy (red) at the $50^{\text{th}}$, $65^{\text{th}}$, and $80^{\text{th}}$ area percentiles.
\label{fig:avg_profile}}
\end{figure}

Figure \ref{fig:avg_profile} shows the average column density profile surrounding each superbubble in both galaxies, again using the $50^{\text{th}}$, $65^{\text{th}}$, and $80^{\text{th}}$ area percentiles. As discussed earlier, the typical H~I column density of NGC 6946 is larger than that of the simulation, especially near the bubble centers. The $80^{\text{th}}$ area percentile for the simulated galaxy is an outlier, since only $2$ superbubbles are found and averaged. 

To calculate the average column density profiles, we selected $100$ annuli with a $20\ \mathrm{pc}$ radial width about each superbubble’s centroid, going from the superbubble centroid out to $2\ \mathrm{kpc}$. As expected, column densities near the centers of the superbubbles are lower than the rest of the profile, which increases to a maximum and slowly flattens out. There can be a slight peak before the average profile becomes flat. This may reflect a buildup of gas being bulldozed outward near the edges of the superbubbles.

\subsection{Superbubble Temperature Analysis} \label{subsec:temp_analysis}

\begin{figure*}
\includegraphics{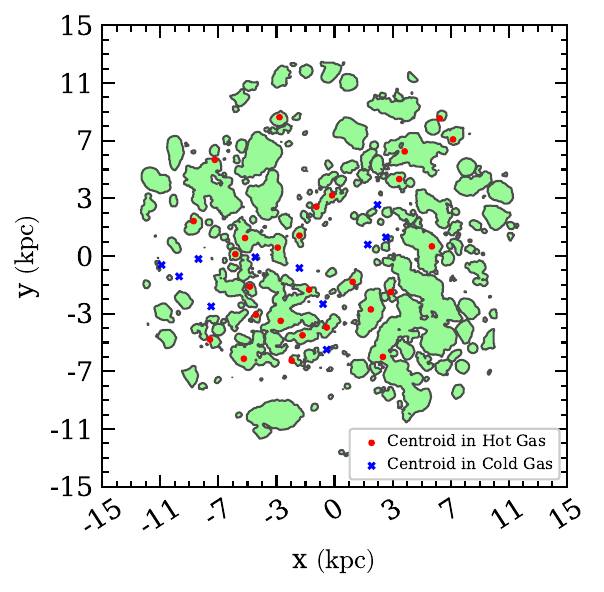}
\caption{Locations of candidate superbubble centroids in the simulated galaxy within a contour map of midplane temperature. A single temperature contour shows where $T=10^5\ \rm K$, and green shading shows where $T > 10^5 \, {\rm K}$. Red circles represent superbubble centroids located in hot gas. Blue crosses represent superbubble centroids located in cold gas. This set of centroids was found using the $50^{\text{th}}$ area percentile threshold, and the galaxy inclination is face-on.
\label{fig:temp_map}}
\end{figure*}

In order to check whether the candidate superbubbles identified in the simulation are real superbubbles, we perform an analysis using temperature information from the simulation. Starting with the superbubbles identified using the $50^{\text{th}}$ area percentile, we look at midplane gas temperatures at the locations of the superbubble centroids.

Figure \ref{fig:temp_map} shows a midplane temperature map with the superbubble centroids plotted as dots or crosses. Here, the galaxy and its superbubbles are treated as if they are not inclined. The majority of superbubble centroids are located in regions of hot gas ($T>10^5\ \rm K$). Those are denoted by red circles. However, some of the superbubble centroids, shown by blue crosses on the plot, are found in regions of cold ($T<10^5\ \rm K$) midplane gas. The prevalence of red dots in the temperature contour map indicates that most of the candidate superbubbles are  associated with regions of hot gas excavated by correlated supernovae.

\begin{figure}
\includegraphics[width=\columnwidth]{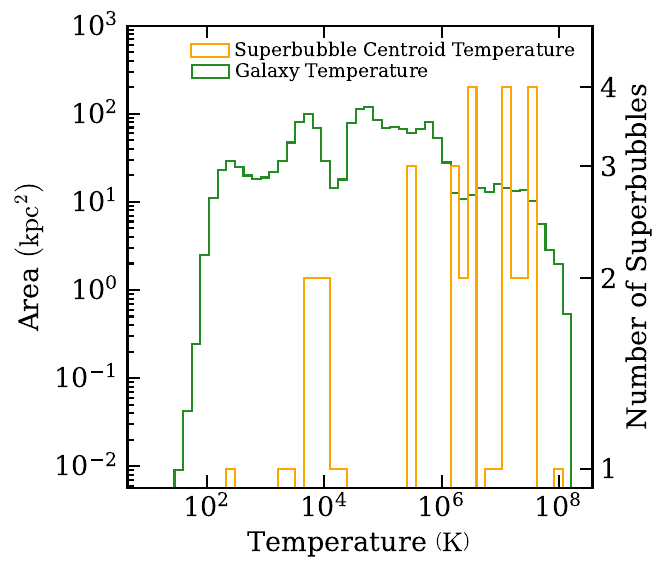}
\caption{Distribution of midplane gas temperature in the simulated galaxy for the superbubble centroids in Figure \ref{fig:temp_map} (orange, right axis) and the entire galactic disk (green, left axis).
\label{fig:temp_hist}}
\end{figure}

We can think of two reasons why some of the candidate superbubble centroids are located in regions of cold gas. The first is that when we identify the superbubbles we do so using the inclined galaxy image shown in Figure \ref{fig:processed_sim}, so we first need to unincline the superbubble centroids to plot them on the midplane temperature slice map. Performing this operation may cause the superbubble centroids to move off of a region of hot gas when plotted on the midplane temperature slice. The second is that the algorithm may combine several separate superbubbles into a single region because of their close proximity to one another. The centroids of the combined superbubbles will then be outside of regions of hot gas because they will be found between the individual superbubbles.

Figure \ref{fig:temp_hist} shows histograms of gas temperature across the entire midplane and at each superbubble centroid to compare their distributions. When comparing them, we see that the temperature distribution at the superbubble centroids is skewed to higher temperatures, confirming that more of the identified superbubbles are found in regions of high temperature. Most of the superbubble centroids are found in gas with temperatures above $10^6$ K, similar to the theoretical values reported in \cite{Weaver_1977}.


\section{Discussion} \label{sec:discuss}

We have developed a superbubble finding algorithm with only one adjustable parameter, the column density percentile threshold $P$, in order to provide a robust and simple superbubble identification method and to minimize the possibility of confirmation bias by avoiding fine-tuning of a large number of parameters. The results of the algorithm are qualitatively insensitive to this value as long as it's between the $50^{\text{th}}$ and $80^{\text{th}}$ area percentiles, as demonstrated in Figure \ref{fig:bubble_persistence}. The consistency of the results when using different area percentiles is a mark of success of our algorithm. Having the same candidate superbubble region being chosen even when using different area percentiles demonstrates the robustness of the algorithm. However, the algorithm does have a few potential drawbacks which we want to discuss. 

We have already mentioned some improvements that we could apply to mitigate some of the known limitations, but we wish to present a few others. We will collect these into three categories: improvements that can be made to the algorithm itself, improvements to the synthetic H~I column density map from the simulated data, and future work that compares simulated galaxies with observed galaxies when using the algorithm.

\subsection{Improvements to the Algorithm} \label{subsec:alg_improvements}

One simplifying assumption we have made is to estimate the area of each superbubble as a perfect circle and find an approximate radius for each superbubble. This means that the algorithm is unable to determine if there are multiple superbubbles that are close together and combined into one region, because we do not measure the shape of the regions and only perform basic size and exterior connectivity masking. We could alter this and use an eccentricity measurement already calculated by the algorithm to find the semi-major and semi-minor axis of each superbubble. \cite{Boomsma_2008} shows an example of a method that does this in their  identification of the superbubbles of NGC 6946. This improvement would preserve some of the shape information of superbubbles when analyzing trends of superbubble properties, e.g., allowing analysis of the distribution of superbubble radii along with their shape properties.

We could also impose additional filters on candidate superbubbles using shape information. We expect superbubbles to be blown out along paths of least resistance in the ISM, meaning they are not necessarily circular as can be seen in Figure \ref{fig:temp_map}, but are expected to be small in scale when compared to the entire galactic disk. By including trends in real superbubble shape, the algorithm would better account for the non-standard shape of superbubbles when identifying candidate regions. With such a restriction, the algorithm may be able to more consistently find superbubbles even with changes in inclination and noise. An example of using improved requirements is shown through the Minkowski maps used in \cite{Collischon_2021} when identifying bubbles. Using shape information could help the algorithm choose superbubbles more robustly, but could also introduce more confirmation bias by allowing for additional tunable parameters.
    
\subsection{Improvements to the Synthetic H~I Map} \label{subsec:synth_HI_improvements}

The conversion from simulated galaxy data to a column density map could be made more realistic by using a spatially variable UV radiation field that is largest at the center of the galaxy and decreases radially outward. We expect this improvement may change the population of identified superbubbles near the centers of galaxies, since the very large UV radiation field in the centers of galaxies can make it more difficult for atomic hydrogen to become molecular (e.g., \citealt{Krumholz_2011}).

Our processing pipeline used to produce a synthetic H~I map is very simple compared to the data processing pipelines used for radio interferometry observations (e.g., \citealt{CASA_2022}). The most significant difference is that our synthetic H~I map does not include the effect of sparse sampling of Fourier modes in the image plane due to a small number of antenna pairs. How this would affect our results is uncertain, but it is conceivable that this may explain the differences in radial profiles of column density apparent in Figure \ref{fig:avg_profile}. Processing the simulation data with the same pipeline used for observations will be necessary in order to draw quantitative conclusions about how well the properties of simulated and observed superbubbles agree with each other.

\subsection{Future Work} \label{subsec:future_work}

Applying our algorithm to a wide range of galaxies, both observed and simulated, will allow us to quantify the variation in superbubble properties across galaxy populations, and will inform our expectations for how closely simulated and observed superbubble properties should agree. We used an H~I map of NGC 6946 for most of our discussion above, but it is possible to apply this algorithm to all of galaxies in the THINGS H~I survey. It will also be of interest to apply our algorithm to galaxies in other H~I surveys in order to empirically test the robustness of our algorithm to differences in noise and beam effects from different radio telescopes and the processing algorithms applied by different radio surveys.

New James Webb Space Telescope (JWST) infrared (IR) dust emission maps can also be explored with the algorithm, as these maps approximately trace H~I and are of high resolution \citep{Williams_2024}. Our algorithm is especially of interest for analyzing superbubbles in IR dust maps, since we only require 2D maps as input, rather than 3D position-position-velocity cubes used by some other H~I superbubble algorithms. The high spatial resolution and low noise of these observations also makes them excellent candidates for comparison to galaxy simulations.

Comparing the catalogues of superbubbles produced by our algorithm from input simulated and observed galaxies tests the realism of the implementation of supernova feedback in simulations in a direct way with only a single adjustable parameter. Because the algorithm was designed to allow for comparison of different galaxies, more analysis of a variety of observed and simulated galaxies could help test a broad range of simulation properties that could be affected by supernova feedback processes. Future work that puts error bars on superbubble properties in both simulations and observations will be necessary in determining the significance in differences between simulated and observed superbubble properties. This work will open the door to robust tests of supernova feedback models in simulations via their direct effects on the observed galactic gas.


\section{Conclusion} \label{sec:conclusion}

We have presented a new algorithm for identifying and measuring superbubbles in H~I column density maps that requires little human input. We compared the candidate superbubbles identified in both a simulated galaxy and an observed galaxy (NGC~6946) to see whether the simulated galaxy created realistic superbubbles. Lessons learned during development of the algorithm motivated use of an area percentile method to find and compare superbubbles in H~I maps of galaxies with different column density distributions.

We found similarities in both size and radial distribution among the candidate superbubbles in the simulated galaxy and NGC 6946, indicating that the simulation may be producing realistic superbubbles. However, even with the similarities in superbubbles, there was no similar trend in superbubble coverage fraction between the simulated galaxy and NGC 6946. We use information about bubble persistence and the average column density profile to make additional comparisons. We are optimistic that applying the algorithm to a large sample of galaxy images, both simulated and observed, will improve its usefulness as a diagnostic tool of feedback models in galaxy simulations and provide a basis for a standardized comparison of superbubbles across observational and theoretical works.

We hope that our work will lead to robust and reproducible tests of stellar feedback models used in simulations of galaxy formation and evolution.


\section*{Acknowledgements}

This research was supported in part by the NSF through grant AAG-2106575.


\section*{Data Availability}

The data that support the findings of this paper are openly available in GitHub at \url{https://github.com/BrockWW/BubbleFindingAlgorithm.git}.


\bibliographystyle{mnras}
\bibliography{mnras_template}


\bsp
\label{lastpage}
\end{document}